\RequirePackage{fix-cm}
\documentclass{iopjournal-arxiv}
\usepackage{amsmath}
\usepackage{amssymb}

\begin{document}

\articletype{Accepted Manuscript}

\title{Cyclic-Quadrature Intracavity Signal Amplification \\for Gravitational-Wave Detectors}

\author{Kaido Suzuki$^{1,2,*}$, Ryo Iden$^1$, Ken-ichi Harada$^3$, and Kentaro Somiya$^{1}$}

\affil{$^1$Department of Physics, Institute of Science Tokyo, Tokyo, Japan}

\affil{$^2$School of Physics, Mathematics and Computing, The University of Western Australia, Perth, Australia}

\affil{$^3$Nanofiber Quantum Technologies, Inc., 1-22-3 Nishiwaseda, Shinjuku, Tokyo 169-0051, Japan}

\affil{$^*$Author to whom any correspondence should be addressed.}

\email{ksuzuki@gw.phys.sci.isct.ac.jp}

\keywords{gravitational-wave detector, quantum noise, optical parametric amplifier, internal squeezing}
\vspace{0.5\baselineskip}
{\small\noindent This is the Accepted Manuscript version of an article accepted for publication in \textit{Classical and Quantum Gravity}. IOP Publishing Ltd is not responsible for any errors or omissions in this version of the manuscript or any version derived from it. The Version of Record is available online at \url{https://doi.org/10.1088/1361-6382/aec408}.\par}

\begin{abstract}
The high-frequency sensitivity of laser-interferometric gravitational-wave detectors is limited by quantum shot noise. Increasing the circulating optical power reduces shot noise, but is constrained by thermal effects and optomechanical instabilities. 
We propose cyclic-quadrature intracavity signal amplification, in which an optical parametric amplifier (OPA) acts as a phase-sensitive amplifier for the gravitational-wave signal quadrature. The OPA is placed in a quadrature-rotation cavity formed by adding a quadrature-rotation mirror on the output side of the main interferometer. With this cavity detuned by \(\pi/4\), the optical quadratures rotate by \(\pi/2\) on each round trip, so a signal generated in the phase quadrature contributes to the readout phase quadrature only after odd-numbered passes through the OPA.
Successive contributions to the readout phase quadrature, which are separated by two round trips, have alternating signs, making this two-round-trip evolution anti-resonant. During each two-round-trip cycle, however, the same field experiences one amplification and one deamplification, so neither vacuum squeezing nor parametric gain accumulates. Despite the destructive interference between signal contributions separated by two round trips, the OPA increases the signal component extracted through the output coupler, thereby improving the signal-to-noise ratio.
When the OPA gain is sufficiently large, the response approaches that of an interferometer with an effective power enhancement of \(1/\tau^2\), where \(\tau\) is the amplitude transmissivity of the quadrature-rotation mirror that forms the amplifier cavity.
We apply the proposed scheme to a current gravitational-wave detector with a near-future upgrade and show that it improves the quantum-noise-limited sensitivity over a broad frequency range extending into the kilohertz band.
\end{abstract}

\section{Introduction}

The direct detection of gravitational waves has opened a new observational window on the universe~\cite{Abbott2016Observation}. 
The current generation of ground-based laser-interferometric detectors has observed gravitational waves mainly from compact binary coalescences~\cite{Aasi2015AdvancedLIGO,Acernese2015AdvancedVirgo,Aso2013KAGRA,Abbott2023GWTC3}. 
While these observations have already provided rich information on black holes and neutron stars, further improvement of detector sensitivity, especially in the kilohertz band, is essential for accessing physics that remains difficult to probe with the present sensitivity. 
Important targets in this band include post-merger remnants of binary neutron-star coalescences and possible signals from core-collapse supernovae, both of which can carry information on dense nuclear matter and extreme astrophysical dynamics~\cite{Sarin2021PostMerger,Ackley2020NEMO}.

At high frequencies, the sensitivity of laser-interferometric gravitational-wave detectors is  limited by quantum shot noise~\cite{Caves1981QuantumNoise,Kimble2001QND}. 
Shot noise can be reduced by increasing the circulating optical power in the interferometer, but this approach is constrained by optical absorption, thermal effects, and radiation-pressure-induced optomechanical instabilities~\cite{Braginsky2001ParametricInstability,Sidles2006OpticalTorques}.
These limitations motivate alternative methods for improving the high-frequency quantum-noise-limited sensitivity without simply increasing the optical power.

Squeezed-vacuum injection has become an established technique for reducing quantum shot noise in gravitational-wave detectors~\cite{Goda2008Squeezed,Tse2019QuantumEnhanced,Acernese2019VirgoSqueezing}. 
The vacuum field is squeezed before being injected into the interferometer through the antisymmetric port. 
In contrast to this external squeezing, internal squeezing places an optical parametric amplifier (OPA) inside the main interferometer~\cite{Korobko2019QuantumExpander,Adya2020InternalSqueezing}. 
Both external and internal squeezing aim to reduce quantum noise by squeezing the vacuum field, and their performance is therefore sensitive to optical losses, including injection, propagation, and detection losses.

The same OPA can also be operated as a phase-sensitive amplifier by exploiting its anti-squeezed quadrature.
If the gravitational-wave signal quadrature is amplified  prior to detection, the relative contribution of the detection loss can be reduced. 
This type of loss mitigation using phase-sensitive amplification has been discussed, for example, in the context of amplified squeezed states~\cite{Kwan2024AmplifiedSqueezedStates}. 
While such an external amplifier reduces the influence of detection loss, it cannot change the signal-to-noise ratio generated inside the interferometer. 
To improve the intrinsic quantum-noise-limited sensitivity, the amplification has to be incorporated into the interferometer.
In conventional intracavity signal amplification with the signal-recycling cavity tuned to resonance or with the signal-extraction cavity tuned to anti-resonance, however, the signal gain accumulates over successive round trips together with the phase delay. Consequently, the signal amplification is limited to low frequencies, and the resulting response is equivalent to that obtained by increasing the reflectivity of the signal-recycling mirror~\cite{Somiya2012}.

One way to overcome this limitation is to modify the optomechanical dynamics. In a detuned signal-recycled interferometer, the optical spring enhances the gravitational-wave response near the optomechanical resonance~\cite{Buonanno2001SignalRecycled,Suzuki2025OpticalSpringSRMI}. An intracavity OPA can shift this resonance to higher frequencies by amplifying the phase-quadrature signal before it returns to the interferometer, improving sensitivity near the shifted resonance~\cite{Somiya2016ParametricSignalAmplification,Somiya2023IntracavitySignalAmplification}. 
Such manipulation of the optical spring using intracavity optical nonlinearities has been demonstrated in several experimental configurations~\cite{Otabe2024KerrOpticalSpring,Zhang2023OpticalSpring,Liu2025OpticalSpring,Suzuki2026ParametricOpticalSpring}.

Another novel approach to avoid the simultaneous accumulation of signal gain and phase delay over successive round trips is bidirectional internal squeezing~\cite{Vermeulen2026BidirectionalInternalSqueezing}. 
In this recently proposed scheme, the OPA acts on two counter-propagating fields inside the signal-extraction cavity. 
The incoming vacuum field is squeezed, while the outgoing signal and the internal vacuum fields are amplified. By balancing the amplification and squeezing so that the net round-trip parametric gain remains unchanged,  this configuration can provide a large shot-noise improvement. 
The achievable sensitivity improvement is nevertheless sensitive to internal loss, owing to the squeezed vacuum entering the interferometer and the resonant enhancement of loss-induced vacuum fluctuations in the signal-extraction cavity.

In this work, we propose a further distinct cyclic-quadrature amplification scheme to avoid this simultaneous accumulation.
The OPA is placed inside a quadrature-rotation cavity formed by adding a quadrature-rotation mirror at the output port of the main interferometer.
The key idea is to set the detuning of this quadrature-rotation cavity to $\phi=\pi/4$.
Under this condition, the optical quadratures rotate by $\pi/2$ over each round trip. 
A quadrature amplified by the OPA in one pass is mapped onto the squeezed quadrature in the next pass, leaving the net round-trip parametric gain unchanged. 
Unlike the other schemes in which the signal-recycling cavity is either resonant or antiresonant, the signal does not undergo resonant buildup in the amplifier cavity, so intracavity loss is not resonantly enhanced together with the signal.

The remainder of this paper is organized as follows. Section~2 reviews the conventional signal-extraction and intracavity OPA schemes, leading to the basic concept of the proposed cyclic-quadrature intracavity amplification (CQA) scheme. Section~3 presents the quantum-noise-limited sensitivity of the proposed scheme and compares it with existing configurations. Finally, Section~4 summarizes the main results.

\section{Various configurations for intracavity signal amplification}\label{sec:concept}
\subsection{Quantum noise of a conventional interferometer without an amplifier}

The quantum noise spectrum of an interferometer with the arm cavities operated at the dark fringe can be derived by taking a ratio of the signal field and the fluctuation field obtained at the anti-symmetric port. By defining the input field vector $\boldsymbol{a}$ and the gravitational-wave strain $h$, the output field vector $\boldsymbol{b}$ is given as 
\begin{align}            \boldsymbol{b}=e^{2i\alpha}\boldsymbol{Ka}+e^{i\alpha}{\cal G}\boldsymbol{h}&,\ \ \alpha=\arctan\!\left[\frac{1+r}{1-r}
\tan\!\left(\frac{\omega L}{c}\right)\right]\\
\boldsymbol{K}=
    \begin{pmatrix}
        1 & 0\\
        -{\cal K} & 1
    \end{pmatrix}&,\ \ \ \ 
    \boldsymbol{h}=
    \begin{pmatrix}
        0\\
        h
    \end{pmatrix},
\end{align}
where \(\alpha\simeq\omega/{\gamma}\) is the round-trip phase-delay parameter, \(\omega\) is the measurement angular frequency, and \(\gamma=(c/L)(1-r)/(1+r)\) is the cavity linewidth, with \(r\) denoting the amplitude reflectivity of the input test mass (ITM).
\(\mathcal{K}=8\Omega_0P_0/[mL^2\omega^2(\gamma^2+\omega^2)]\) is the optomechanical coupling factor, where \(\Omega_0\) is the carrier angular frequency, \(P_0\) is the laser power incident on the beam splitter, \(m\) is the mass of each test mass, and \(L\) is the arm length. \(\mathcal{G}=\sqrt{\Omega_0P_0/[\hbar(\gamma^2+\omega^2)]}\) is the transfer coefficient from the gravitational-wave strain to the output light field. The quantum-noise power spectrum is then given by
\begin{align}
S_h&=\frac{4\hbar}{mL^2\omega^2}\left(\frac{1}{\cal K}+\cal K\right).\label{eq:ShFPMI}
\end{align}
The two terms in parentheses, \(1/\mathcal{K}\) and \(\mathcal{K}\), represent shot noise and quantum radiation-pressure noise, respectively. Both contributions have been characterized in the Advanced LIGO detectors~\cite{Capote2025AdvancedLIGOO4}.

Current gravitational-wave detectors employ a signal-extraction mirror (SEM) with amplitude reflectivity \(r_{\rm s}\) and amplitude transmissivity \(t_{\rm s}\) at their antisymmetric port to broaden the bandwidth, forming a resonant sideband extraction (RSE) configuration~\cite{Mizuno1993RSE}.
Outside the signal-extraction cavity (SEC), \(\boldsymbol{a}_{\rm SE}\) and \(\boldsymbol{b}_{\rm SE}\) denote the field incident on the SEM and the field leaving the SEM toward the detector, respectively.
Inside the SEC, \(\boldsymbol{a}'_{\rm SE}\) and \(\boldsymbol{b}'_{\rm SE}\) denote the fields propagating from the SEM toward the beam splitter and from the beam splitter toward the SEM, respectively.
Accounting for the \(\pi/2\) quadrature rotation associated with the antiresonant tuning of the SEC, the input--output relations are given by
\begin{align}            \boldsymbol{{b'}}_\mathrm{SE}&=e^{2i\alpha}\boldsymbol{R}_\frac{\pi}{2}\boldsymbol{K}\boldsymbol{R}_\frac{\pi}{2}\boldsymbol{{a'}}_\mathrm{SE}+e^{i\alpha}{\cal G}\boldsymbol{R}_\frac{\pi}{2}\boldsymbol{h},\label{eq:IO-SEC}\\
\boldsymbol{{b}}_\mathrm{SE}&=t_{\rm s}\boldsymbol{{b'}}_\mathrm{SE}-r_{\rm s}\boldsymbol{{a}}_\mathrm{SE},\ \ \ 
\boldsymbol{{a'}}_\mathrm{SE}=r_{\rm s}\boldsymbol{{b'}}_\mathrm{SE}+t_{\rm s}\boldsymbol{{a}}_\mathrm{SE}.\label{eq:IO-SEM}
\end{align}
Here, \(\boldsymbol{R}_{\theta}\) denotes the quadrature-rotation matrix for a rotation by an angle \(\theta\).
Solving Eqs.~(\ref{eq:IO-SEC}) and (\ref{eq:IO-SEM}) and expressing the external input and output fields in rotated quadrature bases, we obtain
\begin{align}
    \boldsymbol{b}_{\rm SE}
    &= e^{2i\alpha_{\rm SE}}
    \begin{pmatrix}
        1 & 0\\
        -\mathcal{K}_{\rm SE} & 1
    \end{pmatrix}
    \boldsymbol{a}_{\rm SE}
    + e^{i\alpha_{\rm SE}}\mathcal{G}_{\rm SE}\boldsymbol{h},
    \ \ 
    \alpha_{\rm SE}
    = \arctan\!\left[\frac{1-r_{\rm s}}{1+r_{\rm s}}\tan\alpha\right],
    \label{eq:IO-RSE}\\
    \mathcal{K}_{\rm SE}
    &= \frac{1+r_{\rm s}}{1-r_{\rm s}}
       \frac{8\Omega_0P_0}{mL^2\omega^2(\gamma_{\rm SE}^2+\omega^2)},
    \ \ \ 
    \mathcal{G}_{\rm SE}
    = \sqrt{\frac{1+r_{\rm s}}{1-r_{\rm s}}
       \frac{\Omega_0P_0}{\hbar(\gamma_{\rm SE}^2+\omega^2)}}.
    \label{eq:K_se}
\end{align}
Here, \(\alpha_{\rm SE}\simeq\omega/\gamma_{\rm SE}\) is the phase-delay parameter of the RSE interferometer, and \(\gamma_{\rm SE}=\gamma(1+r_{\rm s})/(1-r_{\rm s})\) is its effective linewidth.
The quantum-noise-limited sensitivity of a conventional RSE interferometer is then given by
\begin{align}
    S^\mathrm{SE}_h&=\frac{4\hbar}{mL^2\omega^2}\left(\frac{1}{\cal K_\mathrm{SE}}+\cal K_\mathrm{SE}\right)\label{eq:S-SE}.
\end{align}
Using the new optomechanical coupling factor $\cal K_\mathrm{SE}$, this expression can be written in the same form as Eq.~\eqref{eq:ShFPMI}.

\subsection{Unidirectional OPA in a tuned recycling cavity}\label{sec:Unidirectional}
\begin{figure}[htbp]
    \centering
    \includegraphics[width=0.8\linewidth]{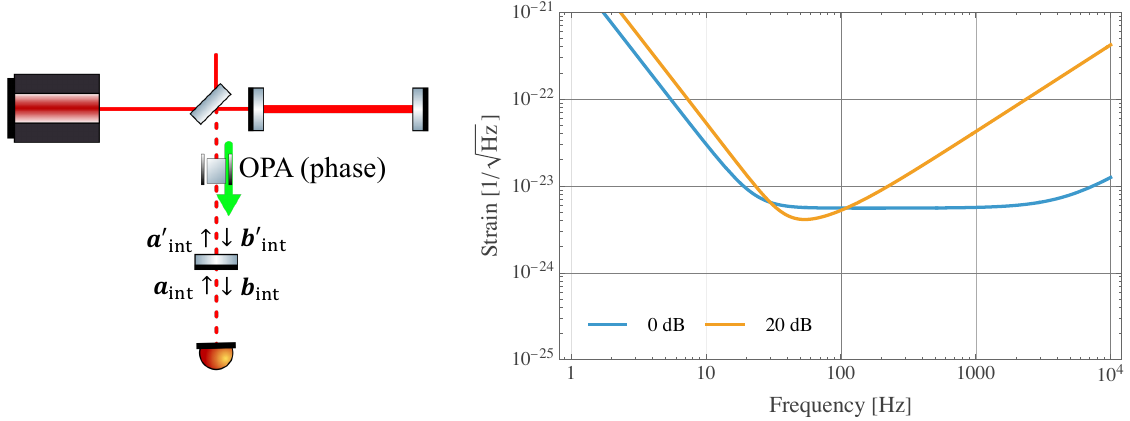}
   \caption{
Unidirectional signal amplification configuration in a tuned signal-extraction cavity.
The orange and blue curves show the quantum-noise limited sensitivity with and without the OPA, respectively.
}
    \label{fig:concept_downstream_sec}
\end{figure}
We first consider placing an OPA inside the SEC to amplify the signal with its pump beam propagating downward, as shown in the left panel of Fig.~\ref{fig:concept_downstream_sec}. The input--output relation in Eq.~(\ref{eq:IO-SEC}) is then modified as follows:
\begin{align}            \boldsymbol{{b'}}_\mathrm{int}&=e^{2i\alpha}\boldsymbol{R}_\frac{\pi}{2}\boldsymbol{SK}\boldsymbol{R}_\frac{\pi}{2}\boldsymbol{{a'}}_\mathrm{int}+e^{i\alpha}{\cal G}\boldsymbol{R}_\frac{\pi}{2}\boldsymbol{Sh},\\
\boldsymbol{{b}}_\mathrm{int}&=t_{\rm s}\boldsymbol{{b'}}_\mathrm{int}-r_{\rm s}\boldsymbol{{a}}_\mathrm{int},\ \ \ 
\boldsymbol{{a'}}_\mathrm{int}=r_{\rm s}\boldsymbol{{b'}}_\mathrm{int}+t_{\rm s}\boldsymbol{{a}}_\mathrm{int}.
\end{align}
where \(\boldsymbol{S}=\operatorname{diag}(1/s,\,s)\) is the squeezing matrix that amplifies the phase quadrature, with \(s\) being the squeezing factor.
The noise spectrum is then given as
\begin{align}
S^\mathrm{int}_h&=\frac{4\hbar}{mL^2\omega^2}\left(\frac{1}{\cal K_\mathrm{int}}+\cal K_\mathrm{int}\right)\label{eq:S-INT},\\
\cal K_\mathrm{int}&
=
\frac{(1- r_{\rm s})^2}{(1- r_{\rm s}/s)^2}\frac{\gamma_{\rm SE}^2+\omega^2}{\gamma_{\rm int}^2+\omega^2}\cal K_\mathrm{SE}
\end{align}
where \(\gamma_{\rm int}\) is the effective linewidth of the RSE interferometer modified by the internal amplifier:
\begin{align}
\gamma_{\rm int}=\frac{c}{L}
\frac{(1-r)(1+r_{\rm s}/s)}
{(1+r)(1-r_{\rm s}/s)}.\label{eq:gammaIFO'}
\end{align}
The right panel of Fig.~\ref{fig:concept_downstream_sec} shows the spectrum without OPA and with a 20\,dB OPA, which corresponds to $s=10$. As can be seen in Eq.~(\ref{eq:S-INT}), the effect of the OPA is equivalent to dividing the reflectivity of the SEM by a factor of \(s\), changing it from \(r_{\rm s}\) to \(r_{\rm s}/s\). 
The OPA amplifies both the real and imaginary parts of the signal field so that the bandwidth of the interferometer decreases with the OPA gain~\cite{Somiya2012}.
Here we assume a mass of \(m=40\,\mathrm{kg}\) for each test mass, an arm length of \(L=4\,\mathrm{km}\), an arm-cavity circulating power of \(P_{\rm arm}=1000\,\mathrm{kW}\), an input-mirror transmissivity of \(t^2=2\%\), and the signal-extraction-mirror transmissivity of \(t_{\rm s}^2=5\%\).

\subsection{Bidirectional intracavity OPA}\label{sec:BiD}
\begin{figure}[htbp]
    \centering
    \includegraphics[width=0.8\linewidth]{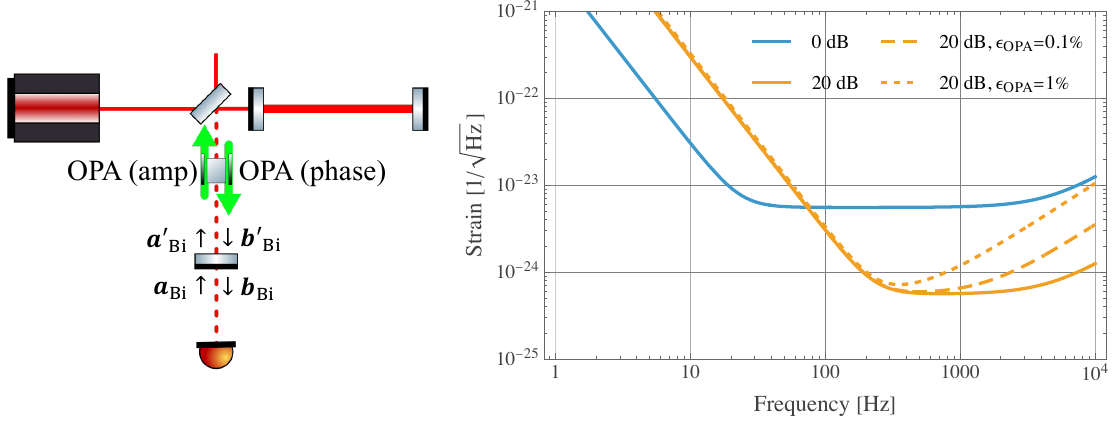}
    \caption{Bidirectional internal squeezing, in which the outgoing signal is amplified while the incoming vacuum is phase-squeezed. The dashed curves show the sensitivity when an optical loss \(\epsilon_{\mathrm{OPA}}\) is included for each pass through the OPA.}
    \label{fig:concept_bidirectional}
\end{figure}
With the same setup using a single-pass OPA inside the tuned SEC but with two pump beams injected upward and downward from the interferometer toward the SEM, one can circumvent the decrease of the bandwidth (see the left panel of Fig.~\ref{fig:concept_bidirectional}). The pump beam propagating downward from the interferometer is set to amplify the phase signal at the OPA, whereas the pump beam propagating upward to the interferometer is set to phase squeeze the vacuum field. Together, these interactions cancel the net parametric gain over one cavity round trip. The input-output relation can be written as
\begin{align}            \boldsymbol{b'}_\mathrm{Bi}&=e^{2i\alpha}\boldsymbol{R}_\frac{\pi}{2}\boldsymbol{SK}\boldsymbol{S}^{-1}\boldsymbol{R}_\frac{\pi}{2}\boldsymbol{a'}_\mathrm{Bi}+e^{i\alpha}{\cal G}\boldsymbol{R}_\frac{\pi}{2}\boldsymbol{Sh},\\
\boldsymbol{{b}}_\mathrm{Bi}&=t_{\rm s}\boldsymbol{{b'}}_\mathrm{Bi}-r_{\rm s}\boldsymbol{{a}}_\mathrm{Bi},\ \ \ 
\boldsymbol{{a'}}_\mathrm{Bi}=r_{\rm s}\boldsymbol{{b'}}_\mathrm{Bi}+t_{\rm s}\boldsymbol{{a}}_\mathrm{Bi}.
\end{align}
The noise spectrum is then given as
\begin{align}
S^\mathrm{Bi}_h&=\frac{4\hbar}{mL^2\omega^2}\left(\frac{1}{\cal K_\mathrm{Bi}}+\cal K_\mathrm{Bi}\right),
\label{eq:S-BiD}\\
\cal K_\mathrm{Bi}&=s^2\cal K_\mathrm{SE},
\end{align}
The resulting quantum-noise sensitivity is equivalent to that of an interferometer operating at an optical power higher by a factor equal to the OPA power gain \(s^2\).
This method provides an elegant solution to the problem of the intracavity OPA scheme discussed in the previous section. 
However, optical losses accumulate together with the signal.
The achievable sensitivity improvement is therefore limited by optical losses inside the SEC, including the excess loss introduced by the OPA. 
In Fig.~\ref{fig:concept_bidirectional}, the OPA loss is parameterized by \(\epsilon_{\mathrm{OPA}}\) and included on each one-way pass through the SEC. 
We use the same interferometer parameters as in Sec.~\ref{sec:Unidirectional}.

\subsection{Intracavity OPA in a detuned interferometer}

A further approach is to place an OPA inside a detuned cavity formed by the SEM and an additional mirror. The gravitational-wave signal in the phase quadrature partially couples to the amplitude quadrature and beats with the carrier field, generating a radiation-pressure force that acts as the restoring force of the {\it optical spring}. By amplifying the phase-quadrature signal, the intracavity OPA actively enhances the optomechanical coupling. The resulting optical-spring resonance frequency is given by
\begin{align}
    \omega_{\rm os}
    \simeq
    \sqrt{
    \frac{8P_0\Omega_0}{mL^2\gamma^2}
    \frac{
    s\sin(2\phi)
    }{
    (r_{\rm s}+1/r_{\rm s})
    -(s+1/s)\cos(2\phi)
    }}.\label{eq:optical_resonance}
\end{align}
Without the OPA (\(s\rightarrow1\)), the first term in the denominator of the second fraction is always larger than the second. With the OPA, however, the denominator can approach zero at a certain value of \(s\), producing a strong optical spring. The gravitational-wave signal response is enhanced near the optical-spring resonance, allowing this technique to possibly improve the high-frequency sensitivity of a gravitational-wave detector. As reported in Ref.~\cite{Somiya2023IntracavitySignalAmplification}, the optical spring eventually disappears from the quantum-noise spectrum as the OPA gain increases, because the noise field amplified inside the cavity exceeds the vacuum-noise contribution that does not enter the cavity.

\begin{figure}[htbp]
    \centering
    \includegraphics[width=0.8\linewidth]{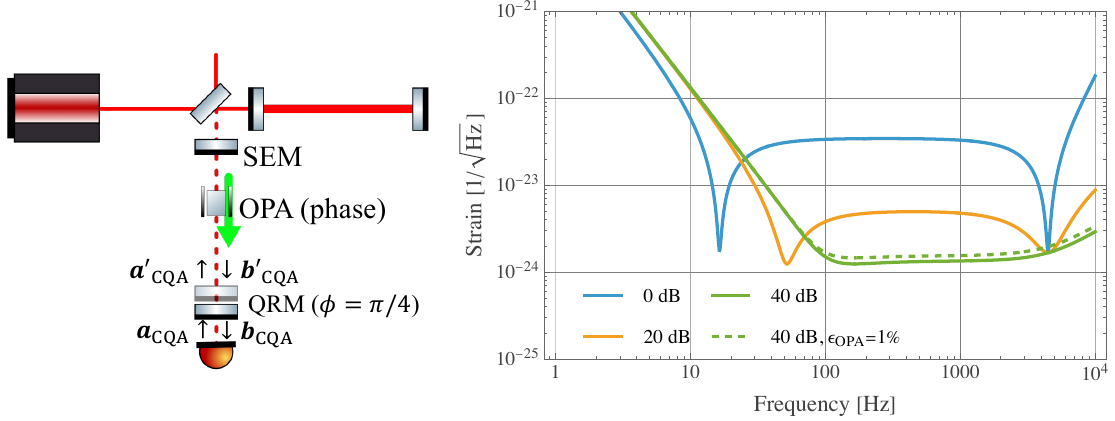}
    \caption{
    Optical configuration of the intracavity OPA in a detuned
    interferometer (left) and the corresponding quantum-noise spectrum (right).
    }
    \label{fig:concept_cqa_rse}
\end{figure}
Figure~\ref{fig:concept_cqa_rse} shows the optical configuration in the left panel and the corresponding quantum-noise spectrum in the right panel. We use the same interferometer parameters as in Sec.~\ref{sec:Unidirectional}. An additional mirror is placed outside the SEC, forming a detuned cavity with the SEM at \(\phi=\pi/4\). 
We refer to this additional mirror as the quadrature-rotation mirror (QRM) and set its power transmissivity to \(5\%\) for this example. The OPA is placed downstream of the SEM inside this cavity and is oriented to amplify the phase quadrature. This configuration is, in principle, optically equivalent to the signal-recycling interferometer proposed in Ref.~\cite{Somiya2023IntracavitySignalAmplification}, while the use of arm cavities allows a higher circulating optical power.

As can be seen from Eq.~\eqref{eq:optical_resonance}, the choice $\phi=\pi/4$ keeps the denominator positive even for a large squeezing factor $s$, allowing the optical spring to be further stiffened as $s$ is increased. Examining the evolution of the optical quadratures under this condition, we find that they rotate by $\pi/2$ over one round trip. Consequently, the quadrature amplified by the OPA on one pass is mapped onto the squeezed quadrature on the next pass. Amplification and squeezing are cyclically exchanged over successive round trips, preventing the round-trip parametric gain from accumulating with the single-pass OPA gain. This basic idea is somewhat analogous to the bidirectional intracavity OPA introduced in Sec.~\ref{sec:BiD}. We refer to this technique as cyclic-quadrature intracavity signal amplification (CQA). In the next section, we discuss the mechanism of this new technique.

\section{Cyclic-quadrature amplification for gravitational-wave detectors}
\subsection{Basic mechanism of cyclic-quadrature amplification}
\label{sec:mechanism}

The scheme was originally intended to improve the sensitivity by exploiting the optical spring. However, focusing on the cyclic cancellation of the OPA gain every two round trips shows that broadband sensitivity enhancement can be achieved without relying on the optical spring, simply by fixing the detuning angle at $\pi/4$. 

As shown in Fig.~\ref{fig:concept_cqa_rse}, an OPA and an additional mirror with amplitude reflectivity \(\rho\) and transmissivity \(\tau\) are placed downstream of the RSE interferometer. The detuning angle of the cavity formed between the SEM and the QRM is fixed at \(\phi=\pi/4\). Using the RSE input--output relation in Eq.~(\ref{eq:IO-RSE}), the input--output relations with the QRM are given by
\begin{align}
\boldsymbol{b}'_{\rm CQA}
&= e^{2i\alpha_{\rm SE}}
\boldsymbol{R}_{\pi/4}\boldsymbol{S}\boldsymbol{K}_{\rm SE}
\boldsymbol{R}_{\pi/4}\boldsymbol{a}'_{\rm CQA}
+ e^{i\alpha_{\rm SE}}\mathcal{G}_{\rm SE}
\boldsymbol{R}_{\pi/4}\boldsymbol{S}\boldsymbol{h},
\quad
\boldsymbol{K}_{\rm SE} =
\begin{pmatrix}
1 & 0\\
-\mathcal{K}_{\rm SE} & 1
\end{pmatrix},
\label{eq:bQR}\\
\boldsymbol{b}_{\rm CQA}
&= \tau\boldsymbol{b}'_{\rm CQA}-\rho\boldsymbol{a}_{\rm CQA},
\qquad
\boldsymbol{a}'_{\rm CQA}
= \rho\boldsymbol{b}'_{\rm CQA}+\tau\boldsymbol{a}_{\rm CQA}.
\label{eq:IO-QRM}
\end{align}
Combining Eqs.~(\ref{eq:bQR}) and (\ref{eq:IO-QRM}), we obtain
\begin{align}
\boldsymbol{b}_{\rm CQA}
&= -\rho\boldsymbol{a}_{\rm CQA}
+ \frac{\tau^2 e^{2i\alpha_{\rm SE}}}
{1-\rho s\mathcal{K}_{\rm SE}e^{2i\alpha_{\rm SE}}
+\rho^2e^{4i\alpha_{\rm SE}}}
\boldsymbol{R}_{\pi/4}
\begin{pmatrix}
1/s & -\rho e^{2i\alpha_{\rm SE}}\\
\rho e^{2i\alpha_{\rm SE}}-s\mathcal{K}_{\rm SE} & s
\end{pmatrix}
\boldsymbol{R}_{\pi/4}\boldsymbol{a}_{\rm CQA}
\nonumber\\
&\quad+
\frac{\tau e^{i\alpha_{\rm SE}}\mathcal{G}_{\rm SE}}
{1-\rho s\mathcal{K}_{\rm SE}e^{2i\alpha_{\rm SE}}
+\rho^2e^{4i\alpha_{\rm SE}}}
\boldsymbol{R}_{\pi/4}
\begin{pmatrix}
-\rho e^{2i\alpha_{\rm SE}}\\
s
\end{pmatrix}
h.
\label{eq:IO-CQA}
\end{align}
The resulting strain-referred quantum-noise spectrum is given by
\begin{align}
S_h^{\rm CQA}
&= \frac{4\hbar}{mL^2\omega^2}
\left[
\frac{1}{\mathcal{K}_{\rm CQA}}+\mathcal{K}_{\rm CQA}
+\frac{4\rho^2}{s^2\tau^4\mathcal{K}_{\rm CQA}}
\cos^2(2\alpha_{\rm SE})
-\frac{4\rho}{s\tau^2}\cos(2\alpha_{\rm SE})
\right],
\label{eq:cqa_sensitivity}\\
\mathcal{K}_{\rm CQA}
&= \frac{\mathcal{K}_{\rm SE}}{\tau^2}.
\end{align}
Here, the readout angle is chosen as \(\zeta=3\pi/4\) so as to detect the OPA-amplified phase-quadrature component after the \(\pi/4\) rotation.
The last two terms in equation~\eqref{eq:cqa_sensitivity} describe the residual finite-gain response. When \(s\) is sufficiently large (\(s\gg1/\tau^2\)), these terms become negligible, and the sensitivity becomes
\begin{align}
    S_h^{\rm CQA}
    \simeq
    \frac{4\hbar}{mL^2\omega^2}
    \left(
        \frac{1}{\mathcal{K_\mathrm{CQA}}}
        +
        \mathcal{K_\mathrm{CQA}}
    \right).
    \label{eq:cqa_power_enhanced_limit}
\end{align}

Equations~(\ref{eq:S-INT}), (\ref{eq:S-BiD}), and (\ref{eq:cqa_power_enhanced_limit}) have the same form as Eq.~(\ref{eq:ShFPMI}), with $\mathcal{K}$ replaced by $\mathcal{K}_{\mathrm{int}}$, $\mathcal{K}_{\mathrm{Bi}}$, and $\mathcal{K}_{\mathrm{CQA}}$, respectively. In particular, $\mathcal{K}_{\mathrm{Bi}}$ and $\mathcal{K}_{\mathrm{CQA}}$ differ from $\mathcal{K}_{\mathrm{SE}}$ only by frequency-independent factors, which means that the sensitivity is improved over a broad frequency band. The cyclic quadrature rotation prevents the accumulation of parametric gain while retaining signal amplification. 

\begin{figure}[htbp]
    \centering
    \includegraphics[width=0.8\linewidth]{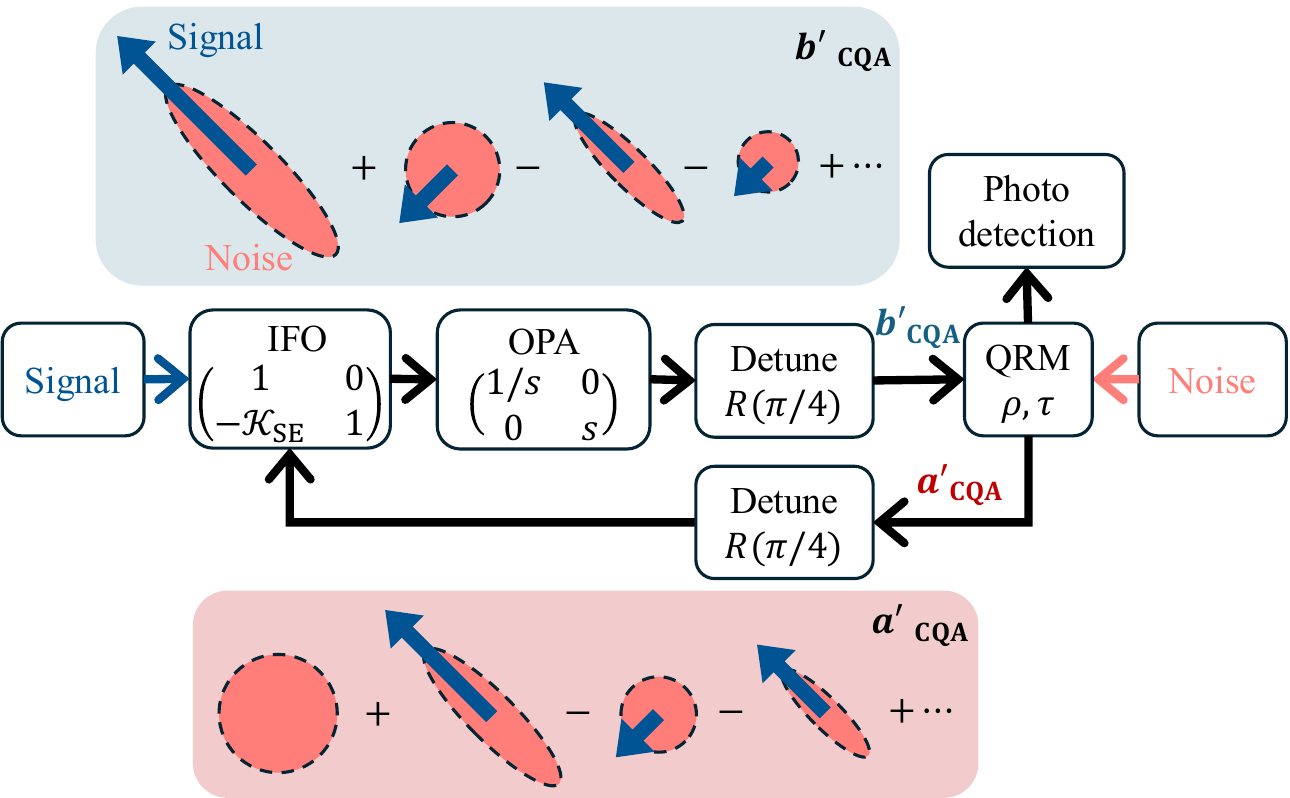}
    \caption{
Schematic of the cyclic-quadrature amplification mechanism. Here, IFO denotes the main interferometer. The upper and lower panels show the successive round-trip contributions to the downstream-propagating field \(\boldsymbol b'_{\rm CQA}\) incident on the QRM and the returning field \(\boldsymbol a'_{\rm CQA}\) propagating toward the interferometer, respectively. The red ellipses represent the quantum-noise uncertainty, while the blue arrows represent the signal displacement.
}
    \label{fig:cqa_mechanismv2}
\end{figure}

Figure~\ref{fig:cqa_mechanismv2} illustrates the field evolution in the cyclic-quadrature configuration. The vacuum field enters through the QRM, while the interferometer generates the gravitational-wave signal in the phase quadrature. After reflection by the detuned QRM, the field returns to the interferometer and encounters the OPA again on the next downstream pass. For \(\phi=\pi/4\), the quadratures rotate by \(\pi/2\) over one round trip. Consequently, the quadrature amplified by \(s\) on one encounter with the OPA is mapped onto the orthogonal quadrature and deamplified by \(1/s\) on the next encounter. The two factors cancel over two successive round trips, preventing the parametric gain from accumulating.

\subsection{Compensation for limited OPA gain}
\label{sec:finite_gain}

The approximation in Eq.~(\ref{eq:cqa_power_enhanced_limit}) is valid when the OPA gain is sufficiently large, i.e. $s\gg 1/\tau^2$. When this condition is not satisfied, the residual finite-gain response produces two characteristic dips in the quantum-noise sensitivity. The high-frequency dip originates from the optical resonance of the signal sidebands, whereas the low-frequency dip is associated with the optomechanical resonance produced by the optical-spring effect~\cite{Somiya2023IntracavitySignalAmplification}. 
The two resonances can be brought together by shifting the detuning from \(\pi/4\) toward resonance. We parameterize the detuning as
\begin{align}
    \phi
    =
    \frac{\pi}{4}-\delta,
\end{align}
where \(\delta>0\) denotes the shift toward resonance. As \(\delta\) is increased, the two sensitivity dips approach each other and merge at a characteristic shift \(\delta_{\rm opt}\), approximately given by
\begin{align}
    \delta_{\rm opt}
    \sim
    \frac{\rho}{s+1/s}.
    \label{eq:optimal_detuning}
\end{align}

\begin{figure}[htbp]
    \centering
    \includegraphics[width=0.5\linewidth]
    {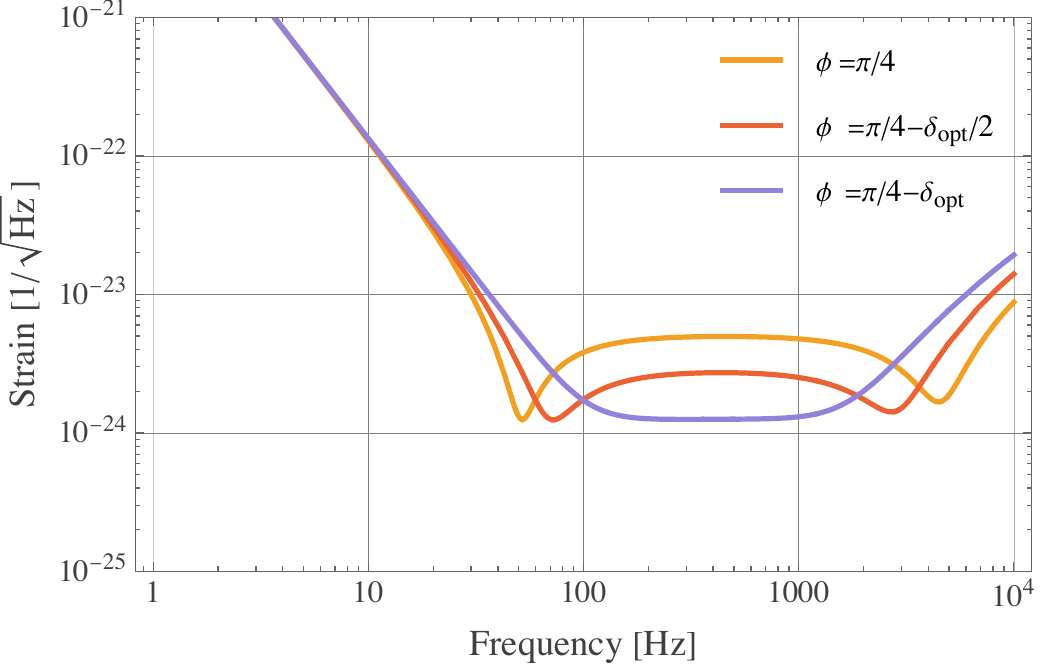}
    \caption{Quantum-noise-limited spectra of an RSE interferometer followed by a cyclic-quadrature amplifier for different detuning phases. The OPA amplitude gain is fixed at \(20\,\mathrm{dB}\), and all other parameters are the same as those used in Fig.~\ref{fig:concept_cqa_rse}.}
    \label{fig:detuning_optimization}
\end{figure}

As shown in Fig.~\ref{fig:detuning_optimization}, the resonant enhancement obtained when the two sensitivity dips merge compensates for the limited OPA gain. Although the resulting sensitivity has a narrower bandwidth, its high-frequency floor can reach the same level as that obtained with a sufficiently large OPA gain.

\subsection{Quantum noise spectrum}
\label{sec:application}

We now estimate the quantum-noise-limited sensitivity achieved with the cyclic-quadrature intracavity amplification technique in a next-generation gravitational-wave detector, using the A+ parameters~\cite{LIGOPostO5}. In what follows, the nominal A+ configuration is referred to as the {\it Baseline} configuration. A modified configuration with a reduced SEM transmissivity, chosen to extend the signal response into the several-kilohertz regime, is referred to as the {\it Wideband} configuration.
This tuning prioritizes sensitivity above approximately \(1\,\mathrm{kHz}\), including the frequency band relevant to binary-neutron-star post-merger signals, at the expense of some mid-frequency sensitivity~\cite{LIGOPostO5}.
These designations are used for both the RSE-only configurations and their corresponding CQA-enhanced configurations. The parameters are listed in Table~\ref{tab:application_parameters}.

\begin{table}[htbp]
    \centering
    \caption{
    Parameters used for the Baseline and Wideband application studies.
    }
    \label{tab:application_parameters}
    \begin{tabular}{lcc}
    \hline
    Parameter
        & Baseline
        & Wideband \\
    \hline
    Carrier wavelength
        & \multicolumn{2}{c}{\(1064\,\mathrm{nm}\)} \\
    Arm length
        & \multicolumn{2}{c}{\(3995\,\mathrm{m}\)} \\
    SEC length
        & \multicolumn{2}{c}{\(55\,\mathrm{m}\)} \\
    CQA cavity length
        & \multicolumn{2}{c}{\(10\,\mathrm{m}\)} \\
    Mass of each test mass
        & \multicolumn{2}{c}{\(40\,\mathrm{kg}\)} \\
    Arm circulating power
        & \multicolumn{2}{c}{\(750\,\mathrm{kW}\)} \\
    ITM power transmissivity
        & \multicolumn{2}{c}{\(1.48\%\)} \\
    SEM power transmissivity \(t_{\rm s}^2\)
        & \(32.5\%\)
        & \(5\%\) \\
    QRM power transmissivity \(\tau^2\)
        & \(50\%\)
        & \(20\%\) \\
    Arm round-trip loss
        & \multicolumn{2}{c}{\(75\,\mathrm{ppm}\)} \\
    SEC round-trip loss
        & \multicolumn{2}{c}{\(3000\,\mathrm{ppm}\)} \\
    Generated external squeezing
        & \multicolumn{2}{c}{\(12\,\mathrm{dB}\)} \\
    Squeezing injection loss
        & \multicolumn{2}{c}{\(5\%\)} \\
    Detection loss
        & \multicolumn{2}{c}{\(4.5\%\)} \\
    CQA-cavity round-trip loss
        & \multicolumn{2}{c}{\(2\%\)} \\
    Internal signal amplifier gain
        & \multicolumn{2}{c}{\(40\,\mathrm{dB}\)} \\
    \hline
    \end{tabular}
\end{table}

The effective OPO loss is set to \(1\%\) per pass, based on the escape efficiency reported in a low-loss squeezing experiment~\cite{Vahlbruch2016SqueezedStates}. Because the field passes through the OPO in both propagation directions, this corresponds to a CQA-cavity round-trip loss of approximately \(2\%\).

The phase-sensitive signal gain inside the CQA cavity is set to \(40\,\mathrm{dB}\), corresponding to an amplitude gain of \(s=100\). This value is assumed as a target for future amplifiers rather than as a performance achievable in the near term. Because CQA directly uses the amplified quadrature for signal amplification, this assumption concerns the internal amplifier gain and does not require the observation of \(40\,\mathrm{dB}\) of squeezing after optical loss.

We compare four configurations. The Baseline and Wideband configurations use SEM transmissivities of \(t_{\rm s}^2=32.5\%\) and \(5\%\), respectively, with the latter chosen to broaden the high-frequency response. The corresponding CQA configurations retain these SEM transmissivities and introduce a QRM with \(\tau^2=50\%\) for the Baseline configuration and \(\tau^2=20\%\) for the Wideband configuration. For both CQA configurations, the detuning is optimized as discussed in Sec.~\ref{sec:finite_gain}.

To evaluate the additional sensitivity improvement provided by CQA in an interferometer that already employs an established quantum-noise-reduction technique, we introduce ideal frequency-dependent external squeezing in all configurations~\cite{Kimble2001QND}. For the conventional RSE configuration, the frequency-dependent squeezing angle is chosen as \(\theta_{\rm RSE}(\omega)=\arctan\mathcal K_{\rm SE}\).
This rotates the injected squeezed field from amplitude squeezing at low frequencies to phase squeezing at high frequencies, thereby suppressing radiation-pressure noise while maintaining shot-noise reduction. In the CQA configuration, the detuned amplifier cavity introduces an additional quadrature rotation, and the squeezing angle is chosen as \(\theta_{\rm CQA}(\omega)=-\pi/4+\arctan\mathcal K_{\rm CQA}\).
\begin{figure}[htbp]
    \centering
    \includegraphics[width=0.6\linewidth]
    {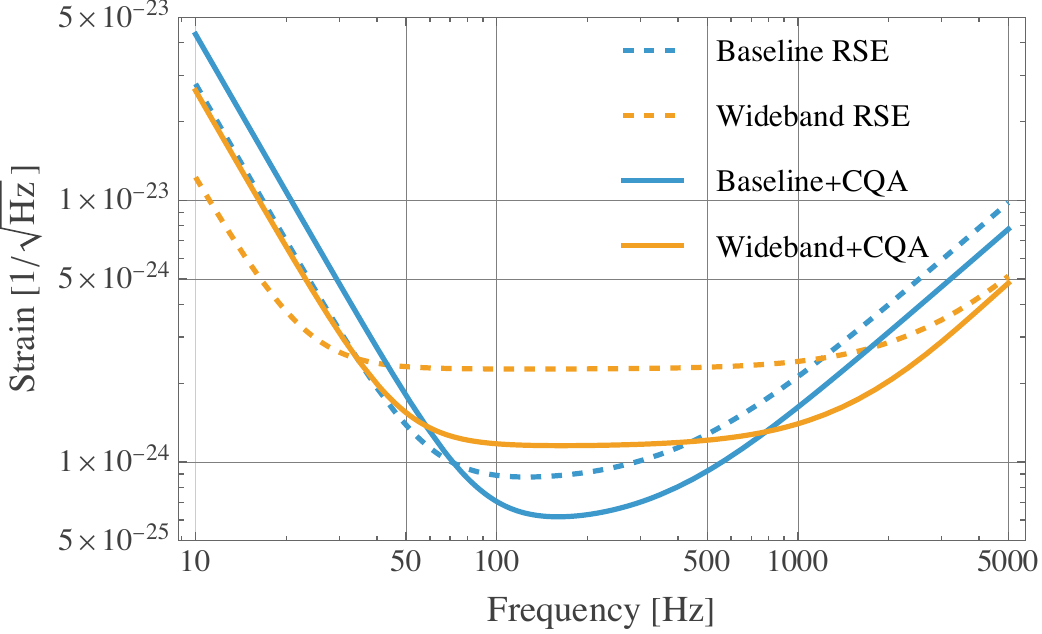}
    \caption{
    Quantum-noise-limited strain sensitivities of the Baseline and Wideband
    configurations, with and without CQA. The parameters are given in
    Table~\ref{tab:application_parameters}.
    }
    \label{fig:cqa_application}
\end{figure}

As shown in Fig.~\ref{fig:cqa_application}, the Baseline RSE+CQA configuration improves the sensitivity over a broad frequency range while largely retaining the overall response of the Baseline RSE configuration.

For the Wideband design, the RSE+CQA configuration improves the quantum-noise-limited sensitivity over a broad frequency range extending into the kilohertz band. Its quantum noise is lower than that of the Wideband RSE configuration over most of the frequency range shown. These results indicate that CQA can extend the high-frequency response of a representative ground-based RSE interferometer beyond that obtained using conventional wideband RSE alone.

\section{Conclusion}
\label{sec:conclusion}
We have proposed a cyclic-quadrature intracavity amplification (CQA) scheme for broadband quantum-noise reduction in interferometric gravitational-wave detectors. With the quadrature-rotation cavity detuned by \(\pi/4\), the optical quadratures rotate by \(\pi/2\) on each round trip, so amplification and deamplification by the intracavity optical parametric amplifier (OPA) cancel over two successive round trips while signal amplification is retained. In the lossless limit with sufficiently large OPA amplitude gain, \(s\gg1/\tau^2\), the quantum-noise spectrum approaches that of the corresponding RSE interferometer with an effective optical-power enhancement of \(1/\tau^2\), where \(\tau\) is the amplitude transmissivity of the quadrature-rotation mirror. This enhancement is achieved without increasing the actual arm-cavity circulating power. The cyclic scheme avoids resonant buildup in the amplifier cavity, so vacuum fluctuations introduced by intracavity loss are not resonantly enhanced.

For finite OPA gain, shifting the cavity detuning from \(\pi/4\) toward resonance allows resonant enhancement to compensate for the limited amplification. The high-frequency noise floor can then reach that of the large-gain configuration, at the expense of a narrower bandwidth.

Quantum-noise calculations using representative A+ detector configurations, including realistic optical losses and frequency-dependent external squeezing, show broadband sensitivity improvements for both the Baseline and Wideband configurations, extending into the kilohertz band.

Beyond gravitational-wave detection, CQA could be explored for interferometric readout of micromechanical displacement and for optically pumped magnetometry. 
In a Michelson--Sagnac interferometer, the displacement signal emerges at the dark port~\cite{Westphal2012Membrane,Sawadsky2015OptomechanicalCoupling}, while in a Faraday-rotation magnetometer, the magnetic-field signal is carried by the polarization mode orthogonal to the probe~\cite{Sierant2026Magnetometry}. 
Coupling these signal modes to a CQA cavity could improve the signal-to-noise ratio at fixed probe power in frequency bands where optical shot noise limits the readout.

\section{Acknowledgments}
This work was supported by Japan Society for the Promotion of Science (23KJ0954, 17H02886, 23K25896), Japan Science and Technology Agency (JPMJSC2209, JPMJCR1873, JPMJAP2320), and Foundation for the Promotion of the Open University of Japan.
\bibliographystyle{iopart-num}
\bibliography{references}
\end{document}